\documentclass[journal]{IEEEtran}

\usepackage{amsthm,amssymb,amsmath,bm}
\usepackage{graphicx}
\usepackage{fancyhdr}
\usepackage{subfigure}
\usepackage[subfigure]{tocloft}
\usepackage[font={small}]{caption}
\usepackage{subfigure}
\usepackage{tabularx}
\usepackage{cite}
\usepackage[Algorithm]{algorithm}
\usepackage{color,soul}

\begin{document}

\title{ Comparison Study between NOMA and SCMA
\thanks{Manuscript received 29 June 2016}
\author{Mohammad.~Moltafet,
Nader.~Mokari, \IEEEmembership{Member, IEEE}, Mohammad~R.~Javan,
\IEEEmembership{Member, IEEE} Paiez.~Azmi, \IEEEmembership{Member,
IEEE}}
%
\thanks{Mohammad Moltafet, Nader Mokari, and Paiez Azmi are with the Department of Electrical Engineering,
Tarbiat Modares University, Tehran, Iran and Mohammad Reza Javan
is with the Department of Electrical Engineering, Shahrood
University of Thechnology, Shahrood, Iran.}} \maketitle
\begin{abstract}
In this paper, the performance and system complexity of the candidate multiple access (MA) techniques for the next generation of cellular systems, namely, non-orthogonal multiple access (NOMA) (in this paper, we consider  power domain MA as NOMA)  and sparse code multiple access (SCMA), are investigated. To this end, for each MA technique, a resource allocation problem considering heterogeneous cellular networks (HetNet)  is formulated. We apply successive convex approximation (SCA) method to each problem and obtain their solutions. The simulation results show that SCMA-based system achieves better performance than NOMA-based one at the cost of more complexity.
\newline
\emph{Index Terms--} NOMA, SCMA, resource allocation, optimization problem, successive convex approximation (SCA).
\end{abstract}

\vspace{-.4 cm}
\section{introduction}
Wireless data traffic is dramatically growing and is expected to grow thousand fold in the next decade \cite{int1,int2}. The fifth generation of wireless networks (5G) is being designed to cope with the excessive data rate demands of future multimedia applications. There are many challenges which should  be addressed in such a network.
Multiple access (MA) techniques have an essential role in improving the performance of mobile communication systems. Non-orthogonal multiple access (NOMA) or power domain MA and sparse code multiple access (SCMA) techniques are promising MA techniques for 5G which have been investigated recently. The main principle of NOMA approach is applying superposition coding (SC) in the transmitter side for assigning each sub-carrier to multiple users and successive interference cancellation (SIC) in the receiver side to cancel the other users signals (interference from other users sharing the same subcarrier). On the other hand, SCMA is a code book based multiple access technique where each subcarrier can be assigned to multiple users with applying an appropriate code book assignment.
In \cite{int3}, the authors studied NOMA performance from
the information theory aspect.  System-level
performance  for NOMA in downlink has been investigated in \cite{int4}. In \cite{int5} and \cite{int6}, the authors evaluated the throughput
and outage of NOMA approach. SCMA as a MA technique has been introduced in \cite{NIC}. In \cite{int7}, a resource allocation method to maximize the energy efficiency in SCMA-based system has been studied. In \cite{int8}, the authors introduced SCMA as a multiple access technique which improves the spectrum efficiency. In \cite{comps}, different NOMA techniques such as pattern division multiple access (PDMA), SCMA, and multi-user shared access
(MUSA) are studied, and their link-level performances are compared with
 each other, in this paper, power domain NOMA is not studied, also, the
 considered MA methods are not investigated from resource allocation and
 receiver complexity perspective. Features, challenges, and future research trend of the MA techniques of 5G have been investigated in \cite{newpap}.

The main contributions of this paper are summarized as
follows:
\begin{itemize}
\item
We consider two MA techniques, namely, NOMA and SCMA, which are candidates of MA techniques in 5G. The performance of these techniques, measured based on the system sum-rate and complexity, is studied and compared to each other.
\item
For each technique, we consider a downlink resource allocation problem in the context of heterogeneous cellular networks (HCN) based on which we evaluate and compare the performance of these techniques.
\item
To solve the resource allocation problems, an iterative algorithm is devised.
In NOMA-based system, in each iteration, power and sub-carriers are allocated separately.
 To solve the subcarrier allocation problem arithmetic geometric mean approximation (AGMA)
 is applied. The power allocation  problem is solved by using SCA approach and applying dual
 method. In SCMA-based system, in each iteration, power and code books are allocated separately
  in which code book allocation is solved by AGMA  and power allocation is solved by using SCA
  approach and applying dual method.
\end{itemize}

This paper is organized as follows. In Section \ref{systemmodelandproblemformulation},
system model and problem formulation for the two MA techniques are presented.
In Section \ref{solutions}, the iterative algorithms to obtain the solution of the
 optimization problems are developed. In Section \ref{complexity}, the implementation
 complexity is investigated. Simulation results are presented in Section \ref{simulations}
  and the paper is concluded in Section \ref{conclusions}.
\vspace{-.2 cm}

\section{system model and problem formulation}\label{systemmodelandproblemformulation}

We consider the downlink of a HCN system with one macro base
station and $F-1$ small base stations (BSs). Total number of users
is $M$ and the total number of available subcarriers is $N$. With
these definitions, the system model and problem formulation of
each MA technique is defined as follows:\vspace{-.2 cm}
\subsection{NOMA-based system}

In this system model, $h_{m,f}^n$ indicates the channel
coefficient between user $m$ and BS $f$ on subcarrier $n$,
$p_{m,f}^n$ shows the transmit power of BS $f$ to user $m$ on
subcarrier $n$ and $\rho_{m,f}^n\in \{0,1\}$ is a binary variable
that indicates the subcarrier allocation of user $m$ in BS $f$,
e.g., $\rho_{m,f}^n=1$ if the subcarrier $n$ is allocated to user
$m$ in BS $f$, otherwise $\rho_{m,f}^n=0$.  Moreover, the set of
BSs is shown by $\mathcal{F}=\{1,2,\dots,F\}$ where MBS  is shown
by $f=1$, the set of all users in BS $f$ is shown by
$\mathcal{M}_f=\{1,2,\dots,M_f\}$ which $\sum_{ f\in
\mathcal{F}}M_f=M$, and the set of all subcarriers is denoted by
$\mathcal{N}=\{1,2,\dots,N\}$.
 Also, for the sake of simplicity the notations $\bold{P}^f_m=[p^f_{m,1},\dots,p^f_{m,N}]^T$, $\bold{P}^f=[p^f_{1},\dots,p^f_{M_f}]^T$, $\bold{P}=[\bold{P}^1,\dots,\bold{P}^F]$, $\boldsymbol{\rho}^f_m=[\rho^f_{m,1},\dots,\rho^f_{m,N}]^T$, $\boldsymbol{\rho}^f=[\rho^f_{1},\dots,\rho^f_{M_f}]^T$ and  $\boldsymbol{\rho}=[\boldsymbol{\rho}^1,\dots,\boldsymbol{\rho}^F]$ are  used.

In NOMA-based systems,  users are sorted based on their channel gain, i.e., $|h^f_{1,n}|^2\ge|h^f_{2,n}|^2\ge\dots\ge|h^f_{M_f,n}|^2$, and for sorted users, we have
$p^f_{1,n}\le p^f_{2,n}\le\dots\le p^f_{M_f,n}$ \cite{int6}. Based on the NOMA approach, in the transmitter side,  users
signals are  multiplexed in power domain by applying  SC, and in the receiver side,  each user removes the other user's signals by using SIC approach. Each user can remove the signals of the users with lower order, and considers the signals of users with higher order as noise. Therefore, the SINR of user $m$ on sub-carrier $n$ in BS $f$ is obtained by
\begin{equation}
\gamma^f_{m,n}=\dfrac{\rho^f_{m,n}|h^f_{m,n}|^2p^f_{m,n}}{I_{m,n}^f+(\sigma^f_{m,n})^2},
\end{equation}
where $I_{m,n}^f$ is obtained as
$I^f_{m,n}= |h^f_{m,n}|^2\sum_{i=1}^{m-1}\rho^f_{i,n}p^f_{i,n}+\sum_{f\in \mathcal{F}/\{f\}}\sum_{{m\in \mathcal{M}_f}} \rho^{f}_{m,n}p^{f}_{m,n}|h^{f}_{m,n}|^2,$
and $(\sigma^f_{m,n})^2$ indicates  the noise power of user $m$ on sub-carrier $n$ in  BS $f$. Therefore, the  rate of user $m$ on subcarrier $n$ in  BS $f$ is obtained by $r^f_{m,n}=\log(1+\gamma^f_{m,n}).$
Accordingly, the system sum-rate is equal to
$\mathbf{R}(\mathbf{P},\boldsymbol{\rho})=\sum_{f\in \mathcal{F}}\sum_{m\in \mathcal{M}_f} \sum_{n\in \mathcal{N}}r^f_{m,n}(\mathbf{P},\boldsymbol{\rho}).$
 Furthermore, we impose a total transmit power constraint for each BS in the system as
$\sum_{m\in \mathcal{M}_f} \sum_{n\in \mathcal{N}}\rho^f_{m,n}p^f_{m,n}\le p^f_{\text{max} }\,\,\,\forall f.$
The proposed resource allocation problem based on NOMA approach  is formulated as:
\begin{subequations}\label{eq8}
\begin{align}\label{eq8a}
&\max_{\boldsymbol{\rho},\mathbf{P}}\; \sum_{f\in \mathcal{F}}\sum_{m\in \mathcal{M}_f} \sum_{n\in \mathcal{N}}r^f_{m,n}(\mathbf{P},\boldsymbol{\rho})\\& \label{eq8b}
 \text{s.t.}:\hspace{.25cm}
\sum_{m\in \mathcal{M}_f} \sum_{n\in \mathcal{N}}\rho^f_{m,n}p^f_{m,n}\le p^f_{\text{max} },\,\,\,\forall f,\\&\label{eq8c}
\hspace{1cm}p^f_{m,n}\ge 0,\,\,\,\forall m,n,f,\\&\label{eq8e}
\hspace{1cm}\sum_{m\in \mathcal{M}_f}\rho^f_{m,n}\le L_T,\,\,\,\forall n,f,\\&\label{eq8f}
\hspace{1cm}\rho^f_{m,n}\in
\begin{Bmatrix}
 0 ,
1
\end{Bmatrix},\,\,\forall m,n,f,
\end{align}
\end{subequations}
where \eqref{eq8e} demonstrates that each subcarrier can be assigned to at most $L_T$ users simultaneously.

\subsection{SCMA-based system}

An SCMA encoder is  a mapping from $\log_2 (J)$ bits to a
$N$-dimensional  codebook of size $J$ \cite{NIC}. The
N-dimensional  codewords of a codebook are sparse vectors with $U$
($U < N$) non-zero entries, which refers to $U$ specific
subcarriers. Based on the SCMA approach, codebooks which are
composed of subcarriers are the basic resource unit in networks
\cite{NIC,int8}, and if each
codebook consists of $U$ subcarriers, there are
$C(N,U)=\dfrac{N!}{(N-U)U!}$ codebooks in the considered system.
The set of codebooks is shown by $\mathcal{C}=\{1,2,\dots,C\}$.
Notation $q^f_{m,c}$ indicates codebook assignment between user
$m$ and codebook $c$  in BS $f$ with $q^f_{m,c}=1$ if codebook $c$
is allocated to user $m$  in BS $f$ and otherwise $q^f_{m,c}=0$.
In addition, notation $\rho^f_{n,c}$  shows the mapping between
subcarriers and codebooks with $\rho^f_{n,c}=1$ if codebook $c$
consists of subcarrier $n$ in BS $f$ and otherwise
$\rho^f_{n,c}=0$. We assume that the mapping between codebooks and
subcarriers are fixed, i.e., $\boldsymbol{\rho}$ is a known
parameter. In addition, notation $p^f_{m,c}$ shows the transmit
power of BS $f$ to user $m$ on codebook $c$. Note that $p^f_{m,c}$
is assigned to subcarrier $n$ in codebook $c$ based on a given
proportion $\eta^f_{n,c}$ with $0\le\eta^f_{n,c}\le 1$ determined
based on codebook design and satisfies $\sum_{\forall n \in
c}\eta^f_{n,c}=1\,\,\forall c$
\cite{NIC,int8}.
 Therefore, the SNR of user $m$ on codebook $c$  in BS $f$ is given by
 \begin{equation}\label{snrscm}
 \gamma^f_{m,c}=\dfrac{q^f_{m,c}\sum_{n\in \mathcal{N}}\eta^f_{n,c}p^f_{m,c}|h^f_{m,n}|^2}{I^f_{m,c}+(\sigma^f_{m,c})^2},
\end{equation}
where $I^f_{m,n}$ is obtained by
$I^f_{m,c}=\sum_{f\in \mathcal{F}/\{f\}}\sum_{{m\in \mathcal{M}_f}}\sum_{n\in \mathcal{N}}q^{f}_{m,c} p^{f}_{m,c}|h^{f}_{m,n}|^2.$
From \eqref{snrscm}, the achievable rate for user $m$ on codebook $c$ is given by
$r^f_{m,c}=\log(1+\gamma^f_{m,c}).$
Accordingly, the system sum-rate is given by
$R_{total}=\sum_{f\in \mathcal{F}}\sum_{{m\in \mathcal{M}_f}}\sum_{c\in \mathcal{C}}r^f_{m,c}(\mathbf{P},\mathbf{Q}).$
 Also, the power constraint for each BS is given by
$\sum_{{m\in \mathcal{M}_f}}\sum_{c\in \mathcal{C}}q^f_{m,c}p^f_{m,c}\le p^f_{\text{max} }.$
 Consequently, the problem formulation of joint power and code book assignment  in SCMA system is formulated as follows:
\begin{subequations}\label{orj_p_sc}
\begin{align}\label{eeq8a}
&\max_{\mathbf{Q},\mathbf{P}}\; \sum_{f\in \mathcal{F}}\sum_{{m\in \mathcal{M}_f}}\sum_{c\in \mathcal{C}}r^f_{m,c}(\mathbf{P},\mathbf{Q})\\& \label{eeq8b}
 \text{s.t.}:\hspace{.25cm}
\sum_{{m\in \mathcal{M}_f}}\sum_{c\in \mathcal{C}}q^f_{m,c}p^f_{m,c}\le p^f_{\text{max} },\,\,\,\forall f,\\&\label{eeq8c}
\hspace{1cm}p^f_{m,c}\ge 0,\,\,\,\forall m,c,f,\\&\label{eeq8e}
\hspace{1cm}\sum_{{m\in \mathcal{M}_f}}\sum_{c\in \mathcal{C}}q^f_{m,c}\rho^f_{n,c}\le K,\,\,\,\forall n,f,\\&\label{eeq8f}
\hspace{1cm}q^f_{m,c}\in
\begin{Bmatrix}
 0 ,
1
\end{Bmatrix},\,\,\forall m,n,f,
\end{align}
\end{subequations}
where \eqref{eeq8e} indicates that each sub-carrier can be reused at most $K$ times.

\section{solution of the proposed problems}\label{solutions}
\subsection{NOMA-based system}
The resource allocation problem of NOMA-based system is non-convex
and includes both integer  and continuous variables. Therefore,
the available methods to solve convex optimization problem can not
be applied directly. To solve this problem, an iterative algorithm
is exploited where in each iteration, the main problem is
decoupled into two sub-problems: subcarrier allocation and power
allocation. In each iteration, the subcarrier allocation is solved
by applying AGMA method. Moreover, the power allocation is
computed by applying SCA for low complexity (SCALE) approach. An
overview of the algorithm to solve the main problem is presented
in Algorithm \ref{table-1}.

\begin{algorithm}
\caption{Overview of the solution algorithm }
\label{table-1}

I: Initialize  $\boldsymbol{\rho}(0)$, $\mathbf{P}(0)$ and set $k=0$ (iteration number).
\\
II: Repeat:
\\
III: Set  $\boldsymbol{\rho}=\boldsymbol{\rho}(k)$ and find a solution for problem \eqref{eq8} by applying SCA approach and
assign it to $\mathbf{P}(k+1)$,\\
IV: Find $\boldsymbol{\rho}(k+1)$ by solving \eqref{eq8}
with  $\mathbf{P}=\mathbf{P}(k+1)$,\\
V: When $||\mathbf{P}(k)-\mathbf{P}(k-1)||\le \Upsilon$
stop. \\Otherwise,\\
 set $k=k+1$ and go back to III.\\
 Output:
 \\
 $\boldsymbol{\rho}(k)$ and $\mathbf{P}(k)$ are adopted for the considered system.
\end{algorithm}

\subsubsection{Sub-carrier allocation}
The problem of sub-carrier allocation is formulated as
\begin{subequations}\label{eeqq8}
\begin{align}\label{equva_po_opt}
&\max_{\boldsymbol{\rho}}\; \sum_{f\in \mathcal{F}}\sum_{m\in \mathcal{M}_f} \sum_{n\in \mathcal{N}}r^f_{m,n}(\boldsymbol{\rho})\\&
 \text{s.t.}:\hspace{.25cm} \eqref{eq8b},\eqref{eq8e},\eqref{eq8f}.
\end{align}
\end{subequations}
To solve problem \eqref{eeqq8}, we relax $\rho^f_{m,n}$  to be a real value between zero and one ($0 \le\rho^f_{m,n}\le 1$). Then, $\rho^f_{m,n}$ can be interpreted as a portion of time that sub-carrier $n$ is assigned to user $m$ in BS $f$ \cite{AGMA1,AGMA2}.
It can be shown that the objective of problem \eqref{eeqq8} can be written as follows:
\begin{align}\label{eq21}
&\min_{\boldsymbol{\rho}}\prod_{f\in \mathcal{F},\atop {m\in \mathcal{M}_f, \atop n\in \mathcal{N}}}(\dfrac{|h^f_{m,n}|^2\sum^{m-1}_{i=1}\rho^f_{i,n}p^f_{i,n}+I^f_{m,n}+(\sigma^f_{m,n})^2}{|h^f_{m,n}|^2\sum^{m}_{i=1}\rho^f_{i,n} p^f_{i,n}+I^f_{m,n}+(\sigma^f_{m,n})^2})\\& \nonumber
 \text{s.t.}:\hspace{.25cm} \eqref{eq8b},\eqref{eq8e},\eqref{eq8f}.
\end{align}
The AGMA inequality is expressed as
$\sum_{i=1}^{K}v_iu_i\ge\prod_{i=1}^{K}v_i^{u_i},$
where $\bold{v}=[v_1,\dots,v_K]$, $\bold{u}=[u_1,\dots,u_K]$ and $\sum_{i=1}^{K}u_i=1$ \cite{AGMA3}.
In order to apply AGMA we  define
$X=|h^f_{m,n}|^2\sum^{m}_{i=1}\rho^f_{i,n}p^f_{i,n}+I^f_{m,n}+(\sigma^f_{m,n})^2.$
By applying AGMA inequality we have
\begin{align}\nonumber
&X\ge \underline{X}=\prod_{f\in \mathcal{F}/\{f\}}\prod_{{m\in\mathcal{M}_f}}\bigg[\dfrac{\rho^{f}_{m,n}p^{f}_{m,n}|h^{f}_{m,n}|^2}{\mathcal{W}^{f}_{m,n}}\bigg]^{\mathcal{W}^{f}_{m,n}} \\&\nonumber\times\prod_{i=1}^{m}\bigg[\dfrac{|h^f_{m,n}|^2\rho^f_{i,n}p^f_{i,n}}{\mathcal{R}^f_{i,n}}\bigg]^{\mathcal{R}^f_{i,n}},
\end{align}
where $\mathcal{W}^{f}_{m,n}=\dfrac{p^{f}_{m,n}|h^{f}_{m,n}|^2}{\underline{X}}$ and $\mathcal{R}^f_{i,n}=\dfrac{|h^f_{m,n}|^2\rho^f_{i,n}p^f_{i,n}}{\underline{X}}$.
Consequently, the subcarrier allocation problem is written as follows:
\begin{align}\label{eq22}
&\min_{\boldsymbol{\rho}}\prod_{f\in \mathcal{F},\atop{m\in \mathcal{M}_f,\atop n\in \mathcal{N}}}(\dfrac{|h^f_{m,n}|^2\sum^{m-1}_{i=1}\rho^f_{i,n}p^f_{i,n}+I^f_{m,n}+(\sigma^f_{m,n})^2}{\underline{X}})\\& \nonumber
 \text{s.t.}:\hspace{.25cm} \eqref{eq8b},\eqref{eq8e},\eqref{eq8f}.
\end{align}
This problem is in geometric programming (GP) form and can be solved by available optimization toolboxes like CVX \cite{cvx}.
\subsubsection{Power allocation}
The problem of power allocation is formulated as follows:
\begin{subequations}\label{eeqqq8}
\begin{align}\label{po_al}
&\max_{\boldsymbol{P}}\; \sum_{f\in \mathcal{F}}\sum_{m\in \mathcal{M}_f} \sum_{n\in \mathcal{N}}\rho^f_{m,n}r^f_{m,n}(\mathbf{P})\\& \nonumber
 \text{s.t.}:\hspace{.25cm} \eqref{eq8b},\eqref{eq8c}.
\end{align}
\end{subequations}

To apply SCALE method, an inequality is used to approximate the objective function with a tight lower bond as follows \cite{SCALE}:
\begin{equation}\label{eq11}
\alpha \,\log(z)+\beta\le \,\log(1+z),
\end{equation}
where
$\alpha=\dfrac{\mathcal{Z}_0}{\mathcal{Z}_0+1},~\beta=\log(1+\mathcal{Z}_0)-\dfrac{\mathcal{Z}_0}{\mathcal{Z}_0+1}\,\log(\mathcal{Z}_0).$
By applying inequality \eqref{eq11}, the objective function of problem \eqref{equva_po_opt} is replaced by
$\sum_{f\in \mathcal{F}}\sum_{m\in \mathcal{M}_f} \sum_{n\in \mathcal{N}}\alpha^f_{m,n}\log(\gamma^f_{m,n})+\beta^f_{m,n}.$
Then by transforming $p^f_{m,n}=\exp(\tilde{p}^f_{m,n})$, the standard form of
 convex maximization problem in the new variables
$\tilde{p}^f_{m,n}$ is achieved as follows:
\begin{align}\label{po_al_e}
&\max_{\tilde{\mathbf{P}}}; \sum_{f\in \mathcal{F}}\sum_{m\in \mathcal{M}_f} \sum_{n\in \mathcal{N}}\alpha^f_{m,n}\log(\gamma^f_{m,n})+\beta^f_{m,n}\\& \nonumber
 \text{s.t.}:\hspace{.25cm}
\sum_{m\in \mathcal{M}_f} \sum_{n\in \mathcal{N}}\rho^f_{m,n}\exp(\tilde{p}^f_{m,n})\le p^f_{\text{max} },\,\,\,\forall f,\\& \nonumber
\hspace{1cm}\exp(\tilde{p}^f_{m,n})\ge 0,\,\,\,\forall m,n,f.\\& \nonumber
\end{align}
To show the concavity of objective function, we rewrite it as follows:
\begin{align}\nonumber
&\sum_{f\in \mathcal{F}}\sum_{m\in \mathcal{M}_f} \sum_{n\in \mathcal{N}}\alpha^f_{m,n}\bigg(\log(\rho^f_{m,n}|h^f_{m,n}|^2)+\tilde{p}^f_{m,n}\\\nonumber& - \log\bigg(|h^f_{m,n}|^2\sum^{m-1}_{i=1}\rho^f_{i,n}\exp(\tilde{p}^f_{i,n})\\\label{concpr}&+\sum_{f\in \mathcal{F}/\{f\}}\sum_{{m\in \mathcal{M}_f}} \rho^{f}_{m,n}\tilde{p}^{f}_{m,n}|h^{f}_{m,n}|^2+N_0\bigg)\bigg)+\beta ^f_{m,n}.
\end{align}
Each term in \eqref{concpr} is concave, and therefore, the new objective function is concave.
 We note that log-sum-exp function is convex \cite{boyd}.

To find a power allocation better than
  $\mathbf{P}(k)$ ($k$ is the outer loop iteration number), when $\boldsymbol{\rho}=\boldsymbol{\rho}(k)$
  an iterative power allocation  algorithm that has been shown in Algorithm \ref{table-2} is used. The algorithm can be started by simple high-SINR approximation
 $\boldsymbol{\beta}^0=0$
 and $\boldsymbol{\alpha}^0=1$.
 The output of this algorithm is $\mathbf{P}(k+1)$, and $\mathbf{P}(k)^s$ ($s$ is the inner loop iteration
 number) is considered as power allocation calculated after $k^{\text{th}}$ iteration.
 Finally, we have
 $\mathbf{P}(k+1)=\mathbf{P}(k)^S$ where $S$ is the maximum predefined inner loop iteration number.
\begin{algorithm}
\caption{POWER ALLOCATION ALGORITHM}
\label{table-2}
I: Initialization: Set $s=0$, $\mathbf{P}(k)^0=\mathbf{P}(k)$, $\boldsymbol{\beta}=0$
 and $\boldsymbol{\alpha}=1$,\\
II: Repeat:
\\
III:  Find $\mathbf{P}(k)^s$,  by solving problem \eqref{po_al_e},\\
IV: Update $\boldsymbol{\beta}$
 and $\boldsymbol{\alpha}$ at $\mathcal{Z}_0=\gamma^n_{m,f}(\mathbf{P}(k)^s)$,\\
V: When $s=S$ or convergence,
stop,\\ otherwise,\\
 set $s=s+1$ and go back to III,\\
 Output:
 $\mathbf{P}(k+1)=\mathbf{P}(k)^S.$
\end{algorithm}
To solve the  convex power allocation  problem, dual method is applied. The Lagrangian function of the problem  \eqref{po_al_e} is formulated as:
\begin{align}\nonumber
& L(\mathbf{\tilde{p}},\boldsymbol{\lambda})=\sum_{f\in \mathcal{F}}\sum_{m\in \mathcal{M}_f} \sum_{n\in \mathcal{N}}\alpha^f_{m,n}\log(\gamma^f_{m,n}(\exp(\tilde{p}^f_{m,n})))+\beta ^f_{m,n}\\\label{dual} &+ \sum_{f\in \mathcal{F}}\, \lambda_f(p^f_{\text{max}} -\sum_{m\in \mathcal{M}_f} \sum_{n\in \mathcal{N}}\rho^n_{m,f}\exp(\tilde{p}^n_{m,f})),
\end{align}
where
$\boldsymbol{\lambda}$ is the vector of  Lagrange multipliers.
The dual objective function is given by
\begin{equation}\label{dfrf}
g(\boldsymbol{\lambda}\,,\,\boldsymbol{\delta})=\max_{\mathbf{\tilde{p}}}L(\mathbf{\tilde{p}},\boldsymbol{\lambda}, \boldsymbol{\delta}).
\end{equation}
The dual  problem is solved by finding stationary point of \eqref{dual}
 with respect to
$\mathbf{\tilde{p}}$ with
$\{ \boldsymbol{\lambda} \}$
fixed. To find the stationary point of \eqref{dfrf}, we write
\begin{align}\label{eq000}
& \dfrac{d(L(\mathbf{\tilde{p}},\boldsymbol{\lambda}, \boldsymbol{\delta})}{d\tilde{p}^f_{m,n}}=0.
\end{align}
Then, after simplifying \eqref{eq000} and  applying transformation $p^f_{m,n}=\exp(\tilde{p}^f_{m,n})$, $p^f_{m,n}$ is achieved as \eqref{mn}.
\begin{figure*}[t]
\begin{align}\label{mn}
p^f_{m,n}=\Bigg[\dfrac{ \alpha^f_{m,n}}{\lambda_f+\sum^{M_f}_{i=m+1}\alpha^n_{i,f}\dfrac{\gamma^f_{i,n}(p^f_{i,n})}{p^f_{i,n}}+\sum_{j\in \mathcal{F}/\{f\}}\sum_{{m\in \mathcal{M}_f}}\alpha^j_{m,n}\dfrac{|h^f_{m,n}|^2\gamma^j_{m,n}}{|h^j_{m,n}|^2p^j_{m,n}}}\Bigg].
\end{align}
\hrule
\end{figure*}
The dual variables are updated by applying the subgradient  method
as follows:
\begin{equation}\label{update}
\lambda^{t+1}_f=[\lambda^{t}_f-\theta(p_{\text{max}}^f-\sum_{m\in
\mathcal{M}_f} \sum_{n\in \mathcal{N}}\rho^f_{m,n}p^f_{m,n})]^+,
\end{equation}
where $[.]^+=\max(.,0)$, $t$ indicates  the iteration number for the sub-problem \eqref{po_al_e}, and $\theta$  is  sufficiently small step-size for updating the dual variables.

An overview of the algorithm is presented in Algorithm \ref{table-3}.
\begin{algorithm}
\caption{ALGORITHM TO FIND STATIONARY POINT }
\label{table-3}
I:  Set $t=0$ and initialize  $\boldsymbol{\lambda}^0_f$,\\
II: Repeat:
\\
III: Compute $\mathbf{p}$ by using \eqref{mn}, \\
IV: Update $\boldsymbol{\lambda}$
  by using \eqref{update}, \\
V: When $||\delta^{t}-\delta^{t-1}||\le \epsilon$
stop.\\ otherwise,\\
 set $t=t+1$ and go back to III.
\end{algorithm}
\vspace{-.3cm}
\subsection{SCMA-based system}

The resource allocation problem \eqref{eeq8a} is also non-convex and contains both integer and continuous  variables. To solve the corresponding resource allocation problem, we go through the same steps as in NOMA-based system. The main problem is solved iteratively. In each iteration, the codebook assignment and power allocation is updated separately, and the iterative algorithm is continued until the convergence. The codebook assignment problem is written as follows:
\begin{subequations}\label{c-b-ass}
\begin{align}
&\max_{\mathbf{Q}}\; \sum_{f\in \mathcal{F}}\sum_{{m\in \mathcal{M}_f}}\sum_{c\in \mathcal{C}}q^f_{m,c}r^f_{m,c}(\mathbf{Q})\\&
 \text{s.t.}:\hspace{.25cm}
\eqref{eeq8b},\eqref{eeq8e}, \eqref{eeq8f}.
\end{align}
\end{subequations}
This problem can be solved by applying the same way adopted for
subcarrier allocation in NOMA-based system. To solve the power
allocation problem, similar to the NOMA-based system, at first,
the SCALE approach is applied. Then the dual method is used. By
applying SCALE approach and transforming
$p^f_{m,c}=\exp(\tilde{p}^f_{m,c})$, the power allocation problem
in standard form of convex problem is achieved by
\begin{align}\label{po_al_sc-e}
&\max_{\tilde{\mathbf{P}}}; \sum_{f\in \mathcal{F}}\sum_{{m\in \mathcal{M}_f}}\sum_{c\in \mathcal{C}}\alpha^f_{m,c}\log(\gamma^f_{m,c})+\beta^f_{m,c}\\& \nonumber
 \text{s.t.}:\hspace{.25cm}
\sum_{{m\in \mathcal{M}_f}}\sum_{c\in \mathcal{C}}q^f_{m,c}\exp(\tilde{p}^f_{m,c})\le p^f_{\text{max} },\,\,\,\forall f,\\& \nonumber
\hspace{1cm}\exp(\tilde{p}^f_{m,c})\ge 0,\,\,\,\forall m,n,f.\\& \nonumber
\end{align}
After applying dual method, takeing derivatives (to find stationary point), and some manipulations,  $p^f_{m,c}$ is given by
\begin{align}\label{mnm}
p^f_{m,c}=\Bigg[\dfrac{ \alpha^f_{m,c}}{\lambda_f+\sum_{j\in \mathcal{F}/\{f\}}\sum_{{m\in \mathcal{M}_f}}\alpha^j_{m,c}\dfrac{\sum_{n\in \mathcal{N}}|h^f_{m,n}|^2\gamma^j_{m,c}}{\sum_{n\in \mathcal{N}}h^j_{m,n}|^2p^j_{m,c}}}\Bigg].
\end{align}

\section{implementation complexity on the receiver side}\label{complexity}

In the NOMA-based system, to achieve appropriate signals in each receiver, SIC method  is applied. The complexity of the SIC method is calculated as follows.

We suppose $G$ sub-carriers is assigned to each user, and in each
subcarrier, there are $L_T$ superimposed signals. We also assume
that in each subcarrier, $L_T-1$ signals (as the interference)
should be canceled. By considering NOMA approach, the received
signal at each receiver is given by
$\bold{y}=\bold{H}\bold{x}+\bold{n},$ where $\bold{H}$ is the
channel matrix of size $G\times L_T$ and
$\bold{x}=(x_1,\dots,x_{L_T})$ is the vector of  transmit signals
of size $L_T$. To estimate the signals, the minimum mean square
error (MMSE) detector is applied. By MMSE, the first estimated
signal is given by $\hat{\bold{x}}=\bold{D}\bold{y}$ where
$\bold{D}$ is the transformation matrix calculated as
$\bold{D}=\min_{\bold{D}}E[\|\bold{x}-\bold{D}\bold{y}\|^2],$
whose solution is given by
$\bold{D}=(\bold{H}^H\bold{H}+\sigma^2\bold{I})^{-1}\bold{H}^H.$
Consequently, the complexity order of SIC receiver is
approximately given by $\mathcal{O}((L_T^3+2L_T^2)(G)(L_T-1))$. We note that the
complexity order of calculating $\bold{A}^{-1}$ and
$\bold{A}^H\bold{A}$ (with size $n\times n$) is $n^3$.

In SCMA-based system on the receiver side, message passing
approach (MPA) method is applied. The complexity order of this
method is given by
$\mathcal{O}(I_T(|\boldsymbol{\pi}|^d)),$\cite{CDMA} where
$\mathbf{\pi}$ indicates  the codebook set size, $I_T$ denotes the
total number of iterations, and $d$ denotes the non-zero elements
in each row of the matrix $\bold{X}$ where
$X=(\bold{x}_1,\dots,\bold{x}_n)$) is the factor graph matrix. In
other word, $d$ is the maximum number of signals superimposed on
each subcarrier. In table \ref{table-5}, with some numerical
examples, we show that the complexity of SCMA receiver is higher
than NOMA receiver.

                 \begin{table}
                 \centering
                 \caption{Comparison between NOMA and SCMA receiver  }
                 \label{table-5}
                 \begin{tabular}{ |c|c|c|c|c|c|c|c|}
                 \hline

                     N & d&U & $L_T$&G& NOMA-complexity & SCMA-complexity\\
                 \hline

                 8 & 3& 2&3&4&360&65856\\

                 \hline

                  10& 4&3&4&5&1920&5184000\\

                 \hline
                 \end{tabular}
                 \end{table}

 As we can see, the complexity of MPA method is  higher than SIC,
 and therefore, the trend of researches is to achieve methods which
  decrease the complexity of SCMA receiver system\cite{int7}.
\vspace{-1cm}

\section{simulation results}\label{simulations}
\begin{figure}
\centering
\includegraphics[width=.35\textwidth]{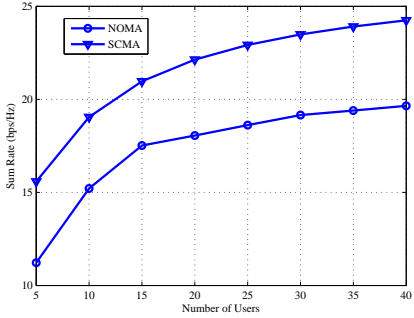}
\caption{Comparison between NOMA and SCMA-based system sumrate versus total number of users.}
\label{pic-1}
\end{figure}

\begin{figure}
\centering
\includegraphics[width=.35\textwidth]{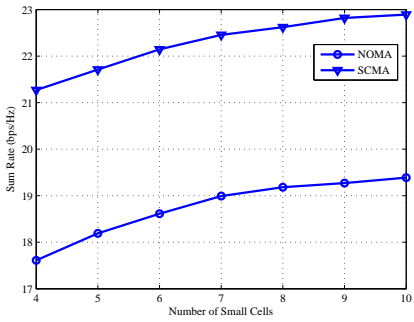}
\caption{Comparison between NOMA and SCMA-based system sumrate versus total number of small cells.}
\label{pic-2}
\end{figure}

In this section, system sum-rate for both system model (NOMA-based
and SCMA-based system) is evaluated under different number of
users and  small cells. In the simulation results,  parameters are
supposed as: the MBS radius is $ .5 $ Km, the SBSs radius are  $
20 $ m, $P^1_{\text{max}} = 10$ (Watts), $P^f_{\text{max}} = 2$
(Watts) for $f\in{2,\dots,F}$, $N = 8$, $\eta^j_{m,c}= 1/2\,\,
\forall f,c,m$, $U=2$, $K=6$, $L_T = 3$, $h^f_{m,n} =
x^f_{m,n}(d^f_{m,n})^{\xi}$ where $\xi$ indicates the path loss
exponent and $\xi = -2$, $x^f_{m,n}$ indicates the Rayleigh
fading, and $d^f_{m,n}$ demonstrates the distance between user $m$
and BS $f$. Figure \ref{pic-1} depicts the system sum-rate versus
the total number of users and Figure \ref{pic-2} depicts the
system sum rate versus number of small cells. As the figures show,
SCMA can achieve larger system sum-rate than NOMA.


\section{conclusion}\label{conclusions}
In this paper, we considered NOMA and SCMA as candidate techniques for multiple access in 5G, and we proposed a resource allocation method for NOMA-based and SCMA-based system to  evaluated the system sum-rate for both of them. Moreover, we calculated the complexity of each system model and compared with each other. The numerical results show,  SCMA technique achieves better system sum rate than NOMA technique, and the system complexity of SCMA is higher than NOMA. Therefore, we can say one of the important challenges of SCMA is designing low complexity receiver.

\end{document}